\documentclass[apj]{emulateapj}
\usepackage{natbib,amsmath} 
\usepackage{CJKutf8}
\usepackage[normalem]{ulem}

\newcommand{\sersic}{S\'{e}rsic}

\newcommand{\Lsun}{$L_{\odot}$}
\newcommand{\Msun}{$M_{\odot}$}
\newcommand{\LIR}{$L_{\rm IR}$}  
\newcommand{\mum}{$\mu m$}
\newcommand{\Mstar}{$M_{*}$}

\begin{document}

\title{The Role of Galaxy Interaction in the SFR-\Mstar\ Relation: Characterizing Morphological Properties of {\it Herschel}-selected Galaxies at $0.2<z<1.5$}

\author{Chao-Ling Hung \begin{CJK*}{UTF8}{gbsn}(洪肇伶)\end{CJK*}\altaffilmark{1}}
%\author{Chao-Ling Hung\altaffilmark{1}}
\author{D. B. Sanders\altaffilmark{1}}
\author{C. M. Casey\altaffilmark{1}}
\author{N. Lee\altaffilmark{1}}
\author{J. E. Barnes\altaffilmark{1}}
\author{P. Capak\altaffilmark{2}}
\author{J. S. Kartaltepe\altaffilmark{3}}
\author{M. Koss\altaffilmark{1}}
\author{K. L. Larson\altaffilmark{1}}
\author{E. Le~Floc'h\altaffilmark{4}}
\author{K. Lockhart\altaffilmark{1}}
\author{A. W. S. Man\altaffilmark{5,1}}
\author{A. W. Mann\altaffilmark{1}}
\author{L. Riguccini\altaffilmark{6,7}}
\author{N. Scoville\altaffilmark{8}}
\author{M. Symeonidis\altaffilmark{9,10}}
%\author{others as appropriate}
\altaffiltext{1} {Institute for Astronomy, University of Hawaii, 2680 Woodlawn Drive, Honolulu, HI 96822, USA; email: clhung@ifa.hawaii.edu}
\altaffiltext{2} {Spitzer Science Center, MS 314-6, California Institute of Technology, Pasadena, CA 91125, USA}
\altaffiltext{3} {National Optical Astronomy Observatory, 950 North Cherry Ave., Tucson, AZ, 85719, USA}
\altaffiltext{4} {UMR AIM (CEA-UP7-CNRS), CEA-Saclay, Orme des Merisiers, b\^at. 709, F-91191 Gif-sur-Yvette Cedex, France}
\altaffiltext{5} {Dark Cosmology Centre, Niels Bohr Institute, University of Copenhagen, Denmark}
\altaffiltext{6} {NASA Ames Research Center, Moffett Field, CA, USA}
\altaffiltext{7} {BAER Institute, Santa Rosa, CA, USA}
\altaffiltext{8} {California Institute of Technology, MC 249-17, 1200 East California Boulevard, Pasadena, CA 91125, USA}
\altaffiltext{9} {University of Sussex, Department of Physics and Astronomy, Pevensey 2 Building, Falmer, Brighton	BN1 9QH, Sussex, UK}
\altaffiltext{10} {University College London, Department of Space \& Climate, Mullard Space Science Laboratory, Holmbury St. Mary, Dorking RH5 6NT, Surrey, UK}

\begin{abstract}

Galaxy interactions/mergers have been shown to dominate the population of infrared luminous galaxies (\LIR\ $\gtrsim10^{11.6}$\Lsun) in the local Universe ($z$\ $ \lesssim 0.25$).
Recent studies based on the relation between galaxies' star formation rates and stellar mass (the SFR-\Mstar\ relation or the ``galaxy main sequence'') have suggested that galaxy interaction/mergers may only become significant when galaxies fall well above the galaxy main sequence.
Since the typical SFR at given \Mstar\ increases with redshift, the existence of galaxy main sequence implies that massive, infrared luminous galaxies at high$-z$ may not necessarily be driven by galaxy interactions.
We examine the role of galaxy interactions in the SFR-\Mstar\ relation by carrying out a morphological analysis of 2084 {\it Herschel}-selected galaxies at $0.2<z<1.5$ in the COSMOS field.   
{\it Herschel}-{\sc PACS} and {\sc -SPIRE} observations covering the full 2-deg$^2$ COSMOS field provide one of the largest far-infrared selected samples of high-redshift galaxies with well-determined redshifts to date, with sufficient sensitivity at $z \sim 1$, to sample objects lying on and above the galaxy main sequence.   
Using a detailed visual classification scheme, we show that the fraction of ``disk-galaxies"  decreases and the fraction of ``irregular" galaxies increases  systematically with increasing \LIR\ out to $z\lesssim1.5$ and $z\lesssim1.0$, respectively.   
At \LIR\ $>10^{11.5}$ \Lsun, $\gtrsim 50$\% of the objects show evident features of strongly interacting/merger systems, where this percentage is similar to the studies of local infrared luminous galaxies. 
The fraction of interacting/merger systems also systematically increases with the deviation from the SFR-\Mstar\ relation, supporting the view that galaxies fall above the main sequence are more dominated by mergers than the main sequence galaxies.
Meanwhile, we find that $\gtrsim18\%$ of massive IR-luminous ``main sequence galaxies" are classified as interacting systems, where this population may not evolve through the evolutionary track predicted by a simple gas exhaustion model.

\end{abstract}

\keywords{galaxies: evolution$-$galaxies: structure$-$infrared: galaxies}

\section{Introduction}
In the hierarchical structure formation paradigm, interactions of galaxies are essential in driving galaxy growth and transforming galaxy morphology \citep{White1978,Barnes1992}.
During violent encounters of galaxies, gas in disk galaxies loses its angular momentum and falls towards the center, inducing enhanced star formation \citep{Barnes1996}.
The UV light emitted from new-born massive stars is scattered and absorbed by dust, and then re-emitted in the far-infrared.
In the local Universe ($z\lesssim0.3$), the majority of Ultraluminous Infrared Galaxies (ULIRGs; \LIR\footnote{\LIR\ $\equiv L_{8-1000\mu m}$ in the object rest-frame}\ $\geq10^{12}$\Lsun) and a significant portion of Luminous Infrared Galaxies (LIRGs; $10^{11}\leq$\LIR$\leq10^{12}$\Lsun) are triggered by major mergers \citep{Sanders1996} of gas-rich disks, as is evident by their morphological features, such as bridges and tidal tails and complex kinematics \citep[e.g.][]{Veilleux2002,Colina2005}. 
These luminous mergers may represent a crucial transition phase from gas-rich disk galaxies to quasi-stellar objects \citep{Sanders1988}. 
Although (U)LIRGs are rare locally, they contribute a significant amount of the star formation rate density at high$-z$ \citep[e.g. $\sim70\%$ at $z\sim1$;][]{Le-Floch2005,Casey2012a}, implying that galaxy interactions play a dominant role in the cosmic star formation history if the origins of local and high$-z$ (U)LIRGs are similar.

Nonetheless, the discovery of a tight correlation between galaxies' star formation rate (SFR) and stellar mass (\Mstar) \citep[also known as the galaxy ``main sequence" (MS);][]{Brinchmann2004,Noeske2007,Elbaz2007} has invoked an alternative picture of galaxy growth incorporating two distinct modes of star formation.
The ``normal star-forming mode'' describes galaxies falling on the SFR-\Mstar\ relation evolving through secular processes such as gas accretion \citep[e.g.][]{Dekel2009, Dave2010a}, while the ``starburst mode'' describes galaxies falling well above the MS which are likely driven by major mergers, representing a star-bursting period with respect to the galaxies on the MS \citep{Rodighiero2011}.
In this scenario, secular accretion in gaseous disks can drive high SFRs in massive galaxies, implying that the majority of (U)LIRGs are not driven by galaxy interactions at high$-z$.
Furthermore, the presence and tightness of the MS over a wide redshift range suggest that star formation histories of galaxies can be simply parametrized based on gas exhaustion/regulation models \citep[e.g.][]{Noeske2007a,Bouche2010}, where major mergers play a minor role on the MS. 

An efficient way to determine the relative importance of galaxy interactions and secular accretion in high$-z$ galaxies is through the characterization of their morphological properties.
For example, high resolution optical images taken with the Advanced Camera for Surveys (ACS) aboard the Hubble Space Telescope ({\it HST}) as part of the Cosmic Evolution Survey \citep[COSMOS;][]{Scoville2007} covering $\sim2$ deg$^2$ can be used to analyze many thousands of galaxies.
In terms of optical morphology, interacting systems often show dramatically different characteristics than isolated disk galaxies.  
Mergers may begin as close pairs without any sign of interaction, then display disrupted structures, tidal features, and then finally coalesce \citep[see][]{Barnes1992}. 
Indeed, these different merger stages are seen out to $z\sim1-2$ in the COSMOS $HST$-ACS images \citep{Kartaltepe2010, Kartaltepe2012}.
Meanwhile, secular accretion from cold gas streams may result in clumpy substructure in gas-rich disks \citep{Dekel2009}, which may explain some of the irregular structures observed in high$-z$ galaxies \citep{Elmegreen2004, Elmegreen2007, Forster-Schreiber2011}.

How do these morphological properties relate to galaxy infrared luminosity, stellar mass, and location on the SFR-\Mstar\ plane?
Among the local (U)LIRGs, the increase in \LIR\ corresponds to the increasing merger fractions of galaxies and the progression of merger stages, from separated galaxy pairs, strongly disturbed interacting systems to more advanced mergers \citep[e.g.][Larson et al. in prep.]{Veilleux2002,Ishida2004,Haan2011}. 
However, it remains unclear if major mergers are also responsible in driving high SFR (e.g. (U)LIRGs) at high$-z$.
\citet{Kartaltepe2010a} show that the fraction of interacting systems systematically increases with \LIR\ at $z\sim1$, which is similar to the trends seen in the local (U)LIRGs.
\citet{Kartaltepe2012} perform a sophisticated visual classification of 52 ULIRGs at $z\sim2$ (with a significant fraction of them lying on the MS), and find that 50\% of them are dominated by interaction/merger systems (73\% including irregular disks).
However, \citet{Wuyts2011a} find that galaxies on the MS have a median \sersic\ index of $n=1$ and the \sersic\ index increases for objects above and below the MS, implying that disk-like galaxies dominate the MS and mergers only dominate those galaxies falling well above the MS.

To understand these results based on different approaches and to examine the role of galaxy interactions in the SFR-\Mstar\ relation, we have carried out a morphological analysis of 2084 {\it Herschel}-selected galaxies at $0.2<z<1.5$ in the COSMOS 2-deg$^2$ field and explored how galaxy morphology varies with \LIR\ and the distance from the MS.
%Identified with the Photodetector Array Camera and Spectrometer (100 and 160 \mum) and the Spectral and Photometric Imaging REceiver (250, 350 and 500 \mum) observations, 
This sample is about twice as large as (three times larger at $1.0<z<1.5$ than) the previous {\it Spitzer} 70 \mum-identified  sample \citep{Kartaltepe2010}, and is also less biased by active galactic nuclei (AGN) since it selects galaxies at longer wavelengths.
We describe the data and sample selection in \S 2 and present our analysis and visual classification scheme in \S 3.
In \S 4, we discuss galaxy morphological properties in three redshift bins.
We discuss the implications of our results in \S 5 and \S 6.
Throughout this paper, we adopt a $\Lambda$CDM cosmology with $H_0=70$ km s$^{-1}$ Mpc$^{-1}$, $\Omega_{M}=0.3$ and $\Omega_{\Lambda}=0.7$ \citep{Hinshaw2009}.
Magnitudes are given in the AB system.

\section{Data and Sample Selection}

The sources studied in this work are identified by the Photodetector Array Camera and Spectrometer \citep[PACS;][]{Poglitsch2010} and the Spectral and Photometric Imaging REceiver \citep[SPIRE;][]{Griffin2010} aboard on the {\it Herschel} Space Observatory \citep{Pilbratt2010}.
The PACS (100 and 160 \mum) and SPIRE (250, 350 and 500 \mum) observations in the COSMOS field are acquired as part of the PACS Evolutionary Probe program \citep[PEP;][]{Lutz2011} and the Herschel Multi-tiered Extragalactic Survey \citep[HerMES;][]{Oliver2012}.
The beamsizes at 100, 160, 250, 350 and 500 \mum\ are 8\arcsec, 12\arcsec, 18\arcsec, 25\arcsec and, 36\arcsec, respectively.

The photometry of {\sc PACS} and {\sc SPIRE} observations are performed at positions of known {\sc MIPS} 24 \mum\ sources \citep{Le-Floch2009} and VLA 1.4 GHz sources \citep{Schinnerer2010}.
This cross-identification source extraction (XID) is described in detail in \citet{Roseboom2010,Roseboom2012}.
A $3\sigma$ detection at 100, 160, 250, 350 and 500 \mum\ corresponds to $\sim$ 5, 10, 8, 11, and 13 mJy, respectively \citep{Roseboom2010,Berta2011}.
The {\it Herschel} photometry measurements are then cross-correlated to their $Ks$-band counterparts using a matching radius of 2\arcsec, and then cross-correlated to the full UV-NIR multiwavelength dataset in the COSMOS field using a matching radius of 1\arcsec \citep[][Lee et al. 2013]{Capak2007,Ilbert2009,McCracken2010}.
Photometric redshifts and stellar masses for $\sim$70\% of the {\it Herschel}-selected detections are thus known from their optical counterparts \citep{Ilbert2009,Ilbert2010}.
Another 20\% of the sample do not have reliable photometric redshift (majority of these sources falls in masked area of the optical images) and the remaining 10 \% are completely obscured in optical images.

Lee et al. 2013 (hereafter L13) measure the infrared luminosity (\LIR), dust temperature and dust mass by fitting the infrared SED of each {\it Herschel}-detected source to a coupled modified grey-body and mid-infrared power-law \citep[method described in][]{Casey2012}.
To ensure secure {\it Herschel} detections that are necessary for accurate SED fitting, L13 restrict the sample to those sources with $\geq$ 5 $\sigma$ detections in at least two out of the five {\it Herschel} {\sc PACS} and {\sc SPIRE} bands and those with photometric redshifts.
They find no apparent selection effects in dust temperature based on these signal-to-noise ratio limits. 
With this restriction, L13 identify a sample of 3457 sources spanning the redshift range $0<z<3.5$ and luminosity range $2\times10^{8}$\Lsun$\leq$\LIR$\leq3.8\times10^{13}$\Lsun.

\section{Analysis}
\subsection{Characterization of {\it Herschel}-selected Galaxies}
We study the morphology of the {\it Herschel}-selected galaxies identified by L13 at $0.2<z<1.5$.
We use the ACS $F814W$ ($I$-band) images taken as part of the COSMOS survey \citep{Koekemoer2007}.
The sample is split into three redshift bins: $0.2<z<0.5$, $0.5<z<1.0$ and $1.0<z<1.5$, where the observed $I$-band corresponds to rest-frame $V$-, $B$-, and $U$-band, respectively.
Given that band-shifting between these three redshift bins may complicate any discussion of morphology evolution, we restrict our discussion of galaxy morphology to each redshift bin separately without addressing evolutionary trends with redshift.
%; for example, the rest-frame UV often traces active star-forming regions which may not necessarily resemble the overall stellar population traced at longer wavelengths.

\begin{figure}
 \centering
  \includegraphics[width=0.5\textwidth]{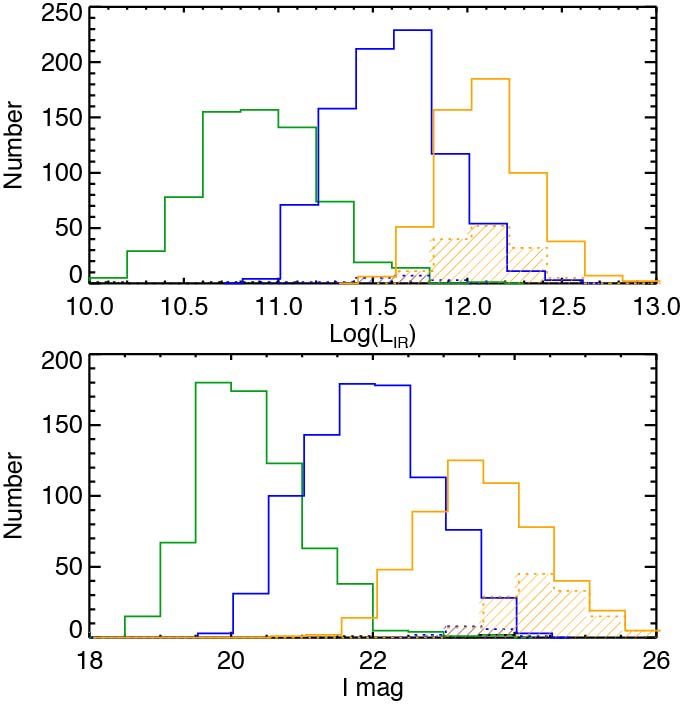} 
\caption{Distribution of {\it Herschel}-selected galaxies included in this study in \LIR\ and $I$ band magnitude. 
Green, blue and orange indicate three redshift bins: $0.2<z<0.5$, $0.5<z<1.0$, and $1.0<z<1.5$.
Shaded area illustrates the distribution of unclassifiable objects in each redshift bin.
Note that although the histograms for different redshift bins have exactly the same location, the distributions are offset slightly from each other in the plot for clarity.} 
\label{fig:lirnum}
\end{figure}

A total of 2084 galaxies are analyzed in this study, with 676, 859 and 549 galaxies at $0.2<z<0.5$, $0.5<z<1.0$ and $1.0<z<1.5$.
Figure~\ref{fig:lirnum} shows the distribution of {\it Herschel}-selected galaxies in \LIR\ and $I$-band magnitude for each redshift bin.
Although our sample contains relatively few infrared luminous galaxies in the lowest redshift bin compared to the other two bins, the data in each redshift bin span a wide range with respect to the SFR-\Mstar\ relation that allows us to probe the morphology on and above the galaxy MS.
The shaded area illustrates the distribution of unclassifiable galaxies (described later in \S 3.2).
The population of unclassifiable galaxies is mostly confined to the highest redshift bin ($1.0<z<1.5$), and these sources are often very faint in ($I$ $>24$). 
Since the fraction of unclassifiable galaxies is relatively small (except for the highest redshift bin), we expect that the unclassifiable galaxies play an minor role in global statistics even if they are biased toward a specific class of sources.

\subsection{Visual Classification Scheme}

\begin{figure}
 %\centering
  \includegraphics[width=0.48\textwidth]{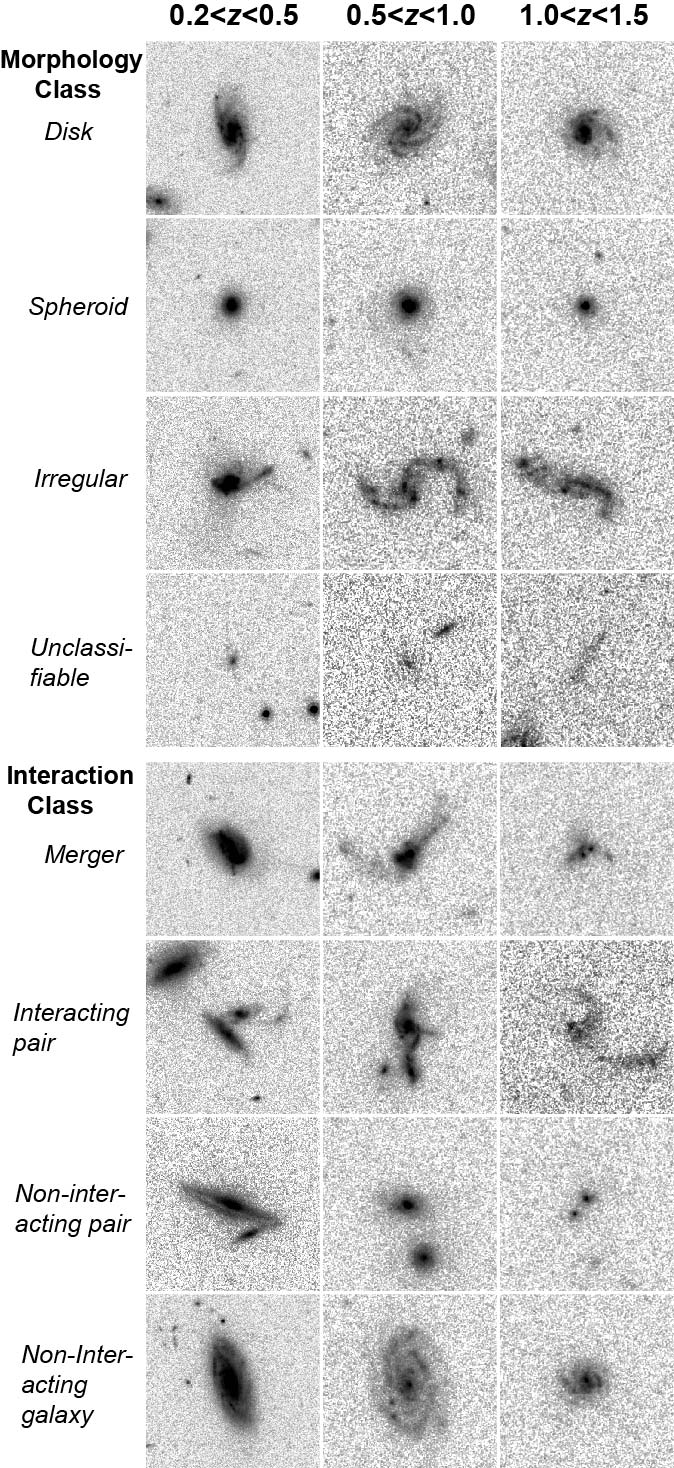} 
\caption{Examples of morphology classes and interaction classes in three redshift bins: $0.2<z<0.5$, $0.5<z<1.0$ and $1.0<z<1.5$. 
Note that the morphology classes provided here are classified as pure disk, pure spheroid and irregular-only galaxies, but in practice galaxies can be classified as having multiple morphology classes.} 
\label{fig:morexample}
\end{figure}

We visually classify the morphological properties of 2084 {\it Herschel}-selected galaxies at $0.2<z<1.5$.
In addition to $F814W$ $I$-band images, we also use the $F125W$ ($J$-band) and $F160W$ ($H$-band) band images taken with the $HST$ Wide Field Camera 3 (WFC3) as part of the CANDELS \citep[Cosmic Assembly Near-infrared Deep Extragalactic Legacy Survey;][]{Grogin2011}.
CANDELS covers $\sim$ 6\% of the COSMOS field, and only 72 galaxies  in the WFC3 coverage meet on {\it Herschel} selection criteria (18 at $0.2<z<0.5$, 37 at $0.5<z<1.0$ and 17 at $1.0<z<1.5$).
The $J$- and $H$- band images allow us to estimate the impact of measuring morphology using different rest-frame wavelengths.
We classify galaxies based on their 50 kpc$\times$50 kpc images, where the corresponding angular sizes of image cutouts are determined from the photometric redshifts of the {\it Herschel} galaxies.
Note that the variation of 50 kpc scale with uncertain photo$-z$ is small at $z>0.2$.

The contrast and brightness schemes used to display the ACS images are critical for identifying morphological features.
We apply two scaling schemes for each image, both of which are used for classifications:
(1) an image displayed with logarithmic scaling and adjustable brightness and contrast;
(2) an image with inverse hyperbolic sine stretch and fixed maximum and minimum values.
The first image is used to demonstrate overall galaxy structure, and the second image is optimized for revealing faint structures.

We adopt the visual classification scheme used by the CANDELS team \citep{Kartaltepe2012,Kocevski2012}, with some minor modifications.
For each galaxy, we identify its morphology class as disk, spheroid, irregular or unclassifiable.
One galaxy can be classified into more than one morphology class such as irregular disks or spheroid with irregular features.
The morphology class has a total of 8 possible outcomes: disk, spheroid, irregular, disk$+$spheroid, disk$+$irregular, spheroid$+$irregular, disk$+$spheroid$+$irregular, and unclassifiable.
In addition to the morphology class, we further assign an interaction class for each galaxy as either merger, interacting pair (major or minor), non-interacting pair (major or minor) or non-interacting galaxy.
Figure~\ref{fig:morexample} provides real examples of each morphology and interaction class in the three redshift bins.
%The use of sub-classes in spheroid and irregular features can minimize the subjectivity in visual classifications.
%In fact, the identifications of prominent spheroid and obvious irregular features are highly consistent between the classification results done by CLH and DBS, and we leave the minor features out of our discussion.

The identification of features associated with galaxy interactions is based on the interaction/mergers systems observed in the local Universe \citep[e.g.][]{Surace1998,Veilleux2002}.
For example, two overlapping disks with disturbed features, obvious single or double nuclei, prominent tidal tails, bridges and loops are all evidence of interactions \citep[e.g.][]{Toomre1972}.
However, the classification of galaxies with ``clumpy structures,'' but no large-scale interacting features, can be ambiguous.
In general, we classify galaxies with symmetric clump distributions as disk-like galaxies.
We introduce an additional ``ambiguous flag'' for galaxies with asymmetric clump distribution; these objects may be candidates for secular-evolving clumpy disks \citep{Elmegreen2004,Elmegreen2007,Dekel2009a}, or alternatively they may also be interacting systems whose large-scale tidal features have faded away. 

\subsection{Robustness of the Visual Classification}

\begin{figure}
 \centering
  \includegraphics[width=0.5\textwidth]{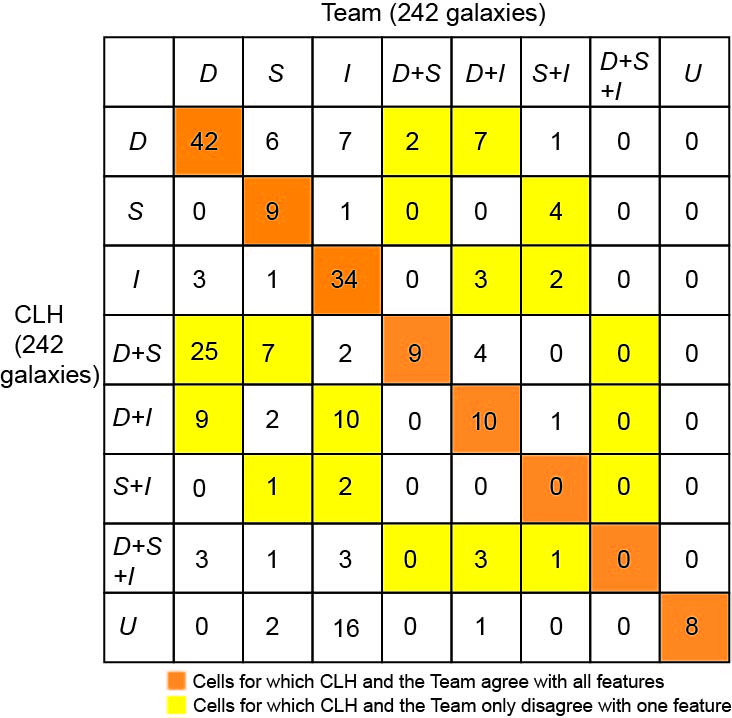} 
\caption{Comparison between CLH and the Team's classification results in a subset of galaxies.
This table categorizes all of eight possible outcomes in the classification of morphology classes, including D (disk only), S (spheroid only), I (irregular only), D+S (disk and spheroid), D+I (disk and irregular), S+I (spheroid and irregular), D+S+I (disk, spheroid, and irregular), and U (unclassifiable).
The table lists the number of galaxies in each cell as belonging to the certain category classified by CLH and the Team.
For illustration, orange cells indicate the categories that CLH and the Team completely agree with all of the morphological features, and yellow cells indicate the categories that only one morphological feature is in disagreement.
Adding the numbers in orange and yellow cells, about 76\% agreement in morphology class is achieved between CLH and the Team.} 
\label{fig:mortable}
\end{figure}

\begin{figure}
 \centering
  \includegraphics[width=0.35\textwidth]{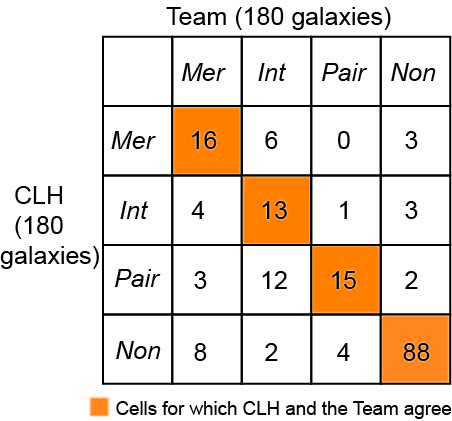} 
\caption{Comparison between CLH and the Team's classification results in a subset of galaxies.
This table categorizes four possible outcomes in the classification of interaction classes, which includes Mer (merger), Int (interacting pair), Pair (non-interacting pair), and Non (non-interacting galaxy).
The table lists the number of galaxies in each cell as belonging to the certain category classified by CLH and the Team.
For illustration,  orange cells indicate complete agreement between CLH and the Team.
Adding the numbers in orange cells, about 75\% agreement in interaction class is achieved between CLH and the Team.} 
\label{fig:inttab}
\end{figure}

\begin{figure}
 %\centering
  \includegraphics[width=0.5\textwidth]{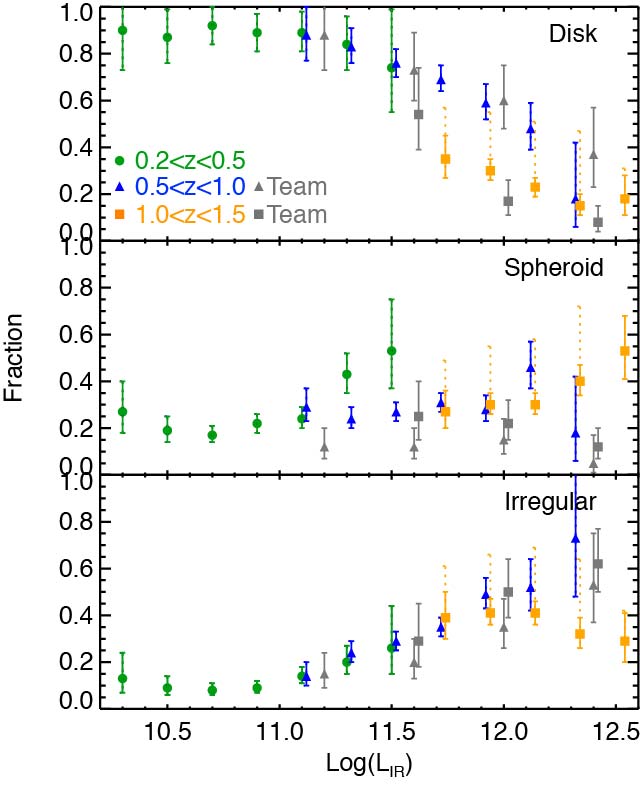} 
\caption{The relation between each morphology class (disk, spheroid and irregular) and \LIR\ in the entire 2084 {\it Herschel}-selected galaxies.
Galaxies in three redshift bins at $0.2<z<0.5$, $0.5<z<1.0$ and $1.0<z<1.5$ are shown as green circle, blue triangle and orange square, respectively.
Gray symbols indicate the classification results of 250 galaxies done by the Team.
The error bars in solid lines are determined assuming a Poisson distribution.
The upper error bars shown as dashed lines represent the fraction of unclassifiable galaxies in each \LIR\ bin (the most conservative estimates of uncertainties).
Note that data points for different redshift bins are offset slightly from each other (in \LIR) for clarity.
 } 
\label{fig:lir_mor}
\end{figure}

Throughout this work, the statistical results of galaxy morphology are based on the classification done by CLH, who perform detailed visual classifications for 2084 {\it Herschel}-selected galaxies.
However, to address the reliability of these classification results based on one classifier, we enlist seven additional classifiers (hereafter the visual classification team, or the Team) to independently classify a subset of 248 galaxies at $0.5<z<1.5$, where this subset is uniformly distributed in \LIR\ but otherwise randomly selected.
To combine the results of the individual classifications in the Team, we only assign morphology and interaction classes to galaxies when more than half of the Team members agree, which results in 242 galaxies in morphology class and 180 galaxies in interaction class.
We compare the classification results between CLH and the Team in this subset in Fig.~\ref{fig:mortable} and~\ref{fig:inttab}.
%Note that since this comparison is done in the earlier stage of this project, both CLH and the Team only use the images with inverse hyperbolic sine scaling for their classifications.
For the morphology class, CLH and the Team completely agree for 46\% of the sample (indicated as orange cells in Fig.~\ref{fig:mortable}). 
The agreement becomes 76\% when including those cases where only one morphological feature is in disagreement (indicated as yellow cells in Fig.~\ref{fig:mortable}).
The agreement for interaction class is 75\% (indicated as orange cells in Fig.~\ref{fig:inttab}).
We compare the statistical results of morphology classes as a function of \LIR\ between CLH (2084 galaxies) and the Team (250 galaxies) in Fig.~\ref{fig:lir_mor}.
In general, the classification of disk and irregular features are consistent, while the spheroid classification yields larger uncertainties.

In the results presented later, we combine morphology and interaction classes into 6 possible outcomes, including merger and major interacting pair, pure spheroid, non-interacting disk, non-interacting pair, minor interacting pair, and unclassifiable galaxy.
We demonstrate our classification by presenting a subset of randomly selected examples in Appendix, and we also examine reliability of our classification results again using this randomly selected subset.

\section{Results}
\subsection{Morphological Properties of $Herschel$-selected Galaxies}
In this section, we examine the fraction of objects represented by each morphology class (disk, spheroid and irregular) as a function of \LIR.
We use a bin size of 0.2 dex in \LIR, which represents the typical uncertainties in \LIR\ measurements. 
Most of the \LIR\ bins contain more than 50 galaxies.
We present the statistical results in three redshift bins.
The $0.5<z<1.0$ bin provides the widest coverage in \LIR\ ($10^{11.0}\lesssim$\LIR$\lesssim10^{12.5}$ \Lsun), whereas the lowest (highest) reshift bin covers the low (high) luminosity end.

Figure~\ref{fig:lir_mor} shows that the fraction of galaxies with disk features systematically decreases with \LIR, and the fraction of those with irregular features systematically increases with \LIR.
These trends are most clearly seen in the two lower redshift bins, $0.2<z<0.5$ and $0.5<z<1.0$.
As \LIR\ increases from $10^{11.0}$ to $10^{12.3}$\Lsun,  the disk fraction decreases from $\sim0.9$ to $\sim0.2$ and the irregular fraction increases from $\sim0.1$ to $\sim0.7$. 
In the highest redshift bin ($1.0<z<1.5$), the low disk fraction at high \LIR\ ($\gtrsim10^{12.2}$\Lsun) and the high irregular fraction at $10^{11.6}\lesssim$\LIR$\lesssim10^{12.2}$ \Lsun\ are consistent with the two lower redshift bins.
However, compared to the two lower redshift bins, the disk fraction in the highest redshift bin is systematically lower at $10^{11.6}\lesssim$\LIR$\lesssim10^{12.2}$ \Lsun, and the irregular fraction is systematically lower at \LIR$\gtrsim10^{12.2}$\Lsun.
The fraction of galaxies with prominent spheroid features shows a slight increasing trend with increasing \LIR\ in all redshift bins.
The spheroid fraction remains 0.2$-$0.3 in most of the \LIR\ bins but increases to $\sim$0.5 at \LIR$\sim10^{12.5}$\Lsun.

The discrepancies in disk and irregular fractions between the two lower redshift bins and the highest one may partly originate from band-shifting, in which the $F814W$ band traces rest-frame $U$ band at $z>1$ and not necessarily resemble the optical morphology.
For example, \citet{Cameron2011} show that $\sim25\%$ of massive galaxies (above $10^{10} M_{\odot}$) at $1.5<z<2.15$ are regular disks at their rest-frame optical wavelengths despite their highly irregular features at UV. 
The fainter UV light and severe surface brightness dimming further desensitize visual classifications at $z>1$, leading to $\sim30\%$  of unclassifiable galaxies.

We have compared our results with the galaxy morphology study which is based on a \textit{Spitzer} 70 \mum-selected sample in the COSMOS field \citep[][hereafter K10]{Kartaltepe2010a}.
The \textit{Spitzer} 70 \mum\ observations identified 1503 galaxies spanning the redshift range $0.01<z<3.5$, with a median redshift of 0.5.
{\it Herschel} observations have allowed us to enlarge the sample of IR luminous galaxies by a factor of two at $0.5<z<1.0$ and by a factor of three at $1.0<z<1.5$.
In K10, the fraction of Spirals among all 1503 galaxies shows a clear decreasing trend with \LIR\ (Fig. 4 in K10), which is consistent with the disk fraction seen in our {\it Herschel}-selected sample in the two lower redshift bins.
We note that $\sim80\%$ of the {\it Spitzer} 70 \mum-selected galaxies are at $z<1$, thus the trends presented in the entire sample mostly reflect the sample at $z<1$.
As for the comparison with spheroid features in the {\it Herschel}-selected galaxies, a slight increasing trend with \LIR\ is also seen in K10 (the sum of Ellipticals and QSOs).
Overall, the correlations between morphological class and \LIR\ are consistent between the current {\it Herschel}-selected galaxies and the previous {\it Spitzer} 70\mum-selected galaxies.

\subsection{The Distribution of Interacting Systems}
%The main goal of this work is to study the role of galaxy interactions in galactic star formation and galaxy assembly.
Here we examine the fraction of interacting systems as a function of \LIR\ and relative offset from the SFR-\Mstar\ ``main sequence'' relation.
For each galaxy, we use the redshift-dependent SFR-\Mstar\ relation corresponding to the galaxy redshift. 
We adopt a functional form of the SFR-\Mstar\ relation from \citet{Bouche2010}:
\begin{equation}
{\rm SFR}=150\ M_{*,11}^{p}(1+z)_{3.2}^{q}\ M_{\odot} yr^{-1},
\end{equation}
where $M_{*,11}\equiv M_{*}/10^{11} M_{\odot}$, $(1+z)_{3.2}\equiv(1+z)/3.2$, $p\simeq0.8$, and $q\simeq2.7$ in the redshift range of $z=0-2$.
To relate \LIR\ to the SFR-\Mstar\ relation, we convert \LIR\ to SFR through:
%\begin{align}
%SFR_{UV+IR} [M_{\odot} yr^{-1}] =&1.09 \times 10^{-10} (L_{IR} + 3.3 L_{2800})/L_{\odot} \nonumber \\
%&\simeq 1.09\times 10^{-10}(L_{IR})/L_{\odot},
%\end{align}
\begin{equation}
{\rm SFR_{ UV+IR}} [M_{\odot} yr^{-1}] \simeq {\rm SFR_{IR}} = 1.09\times 10^{-10}(L_{{\rm IR}}/L_{\odot}),
\end{equation}
\citep{Kennicutt1998,Wuyts2011}, where ${\rm SFR_{IR}}$ represents the reprocessed light from new-born stars, and ${\rm SFR_{UV}}$ represents the unobscured UV light.
In this conversion, we assume that unobscured UV light is negligible in these {\it Herschel}-identified, dusty galaxies \citep{Howell2010}.
Also, we do not account for any AGN contribution, which may be significant at the order of 25\% at \LIR\ $>10^{12}$\Lsun \citep{Veilleux1995,Koss2013}.

\begin{figure}
 %\centering
  \includegraphics[width=0.5\textwidth]{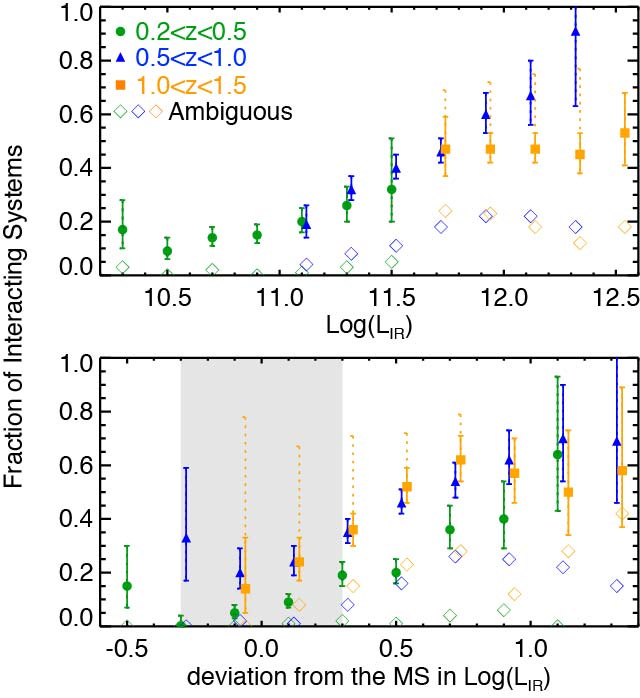} 
\caption{The fraction of interacting systems as a function of \LIR\ and the deviation from the SFR-\Mstar\ relation.
As in Fig.~\ref{fig:lir_mor}, galaxies in three redshift bins at $0.2<z<0.5$, $0.5<z<1.0$ and $1.0<z<1.5$ are shown in green circles, blue triangles and orange squares respectively.
The gray area in the bottom panel illustrates the location of galaxy main sequence with a typical dispersion of 0.3 dex.
The error bars in solid lines are determined assuming a Poisson distribution.
The upper error bars shown as dashed lines represent the fraction of unclassifiable galaxies in each \LIR\ bin (the most conservative estimates of uncertainties).
The diamonds represent the ambiguous galaxies that are identified as interacting systems due to their irregular and clumpy structure without the presence of large-scale tidal features.
Note that data points for different redshift bins are offest slightly from each other (in \LIR) for clarity.
} 
\label{fig:int_three}
\end{figure}

Figure~\ref{fig:int_three} shows the fraction of interacting systems (the sum of mergers and major interacting pairs) as a function of \LIR.
The fraction of interacting systems shows a clear increasing trend with \LIR\ in the two lower redshift bins, and it reaches $\sim0.5$ at \LIR\ $>10^{11.5}$\Lsun.
Our results show the same trend as the ``Major Mergers'' in K10, with a rapid increase at  \LIR\ $\sim10^{11.5}$\Lsun.

To examine the role of galaxy interactions in the SFR-\Mstar\ relation, we further investigate the dependence of the interaction fraction with distance from the SFR-\Mstar\ relation, measured along the \LIR\ axis. 
The {\it Herschel}-selected sample spans from 0.2 dex below to 1.2 dex above the SFR-\Mstar\ relation in each redshift bin.
Given a typical dispersion of 0.3 dex in the SFR-\Mstar\ relation \citep{Noeske2007}, this sample covers galaxies falling on the main sequence through $4\sigma$ outliers.
The bottom panel of Fig.~\ref{fig:int_three} shows an increasing trend that is similar to the trend with \LIR, where the fraction of interacting galaxies increases from $\sim0.2$ on the main sequence to $\sim0.5$ as the deviation achieves $\sim0.5$ dex. 
Taking the $0.5<z<1.0$ bin, the correlation between the interaction fraction and \LIR\ is only slightly stronger than the correlation with the distance from the SFR-\Mstar\ relation (correlation coefficient of 0.98 versus 0.93 with standard errors of 0.09 and 0.13).
Such similar trends are expected since the distance from the SFR-\Mstar\ relation is a function of \LIR.
%The lowest redshift bin shows lower fractions in each bin of \Mstar\ and the deviation from the MS, which is a result of their lower \LIR\ in the low reshift sample.
We further calculate the percentage of interacting systems along the galaxy MS by including galaxies falling within the typical dispersion of 0.3 dex in the SFR-\Mstar\ relation \citep[e.g.][]{Noeske2007}.
At $0.2<z<1.5$, $\sim18\%$ ($\pm2\%$) of the massive ($10^{10}$\Msun$<$\Mstar$<10^{11}$\Msun) infrared luminous galaxies falling on the MS are classified as interacting systems.
In comparison, $\sim51\%$ ($\pm3\%$) of the MS galaxies are classified as non-interacting disks.
The percentage of interacting MS galaxies are larger in the two higher redshift bins ($\sim27\%\pm4\%$ and $\sim26\%\pm5\%$) compared to the lowest redshift bin ($\sim8\%\pm2\%$).
%The systematically lower interaction fraction on and above the MS at $0.2<z<0.5$ may be due to the lower \LIR\ in this redshift bin, which suggests that the correlation between the interaction fraction and the distance above the SFR-\Mstar\ relation is dominated by the correlation with \LIR.
%However, since the interaction fraction is derived based on different rest-frame morphology in three redshift bins, band-shifting may also be a possible cause of such discrepancy. 

\begin{figure*}
 \centering
  \includegraphics[width=0.9\textwidth]{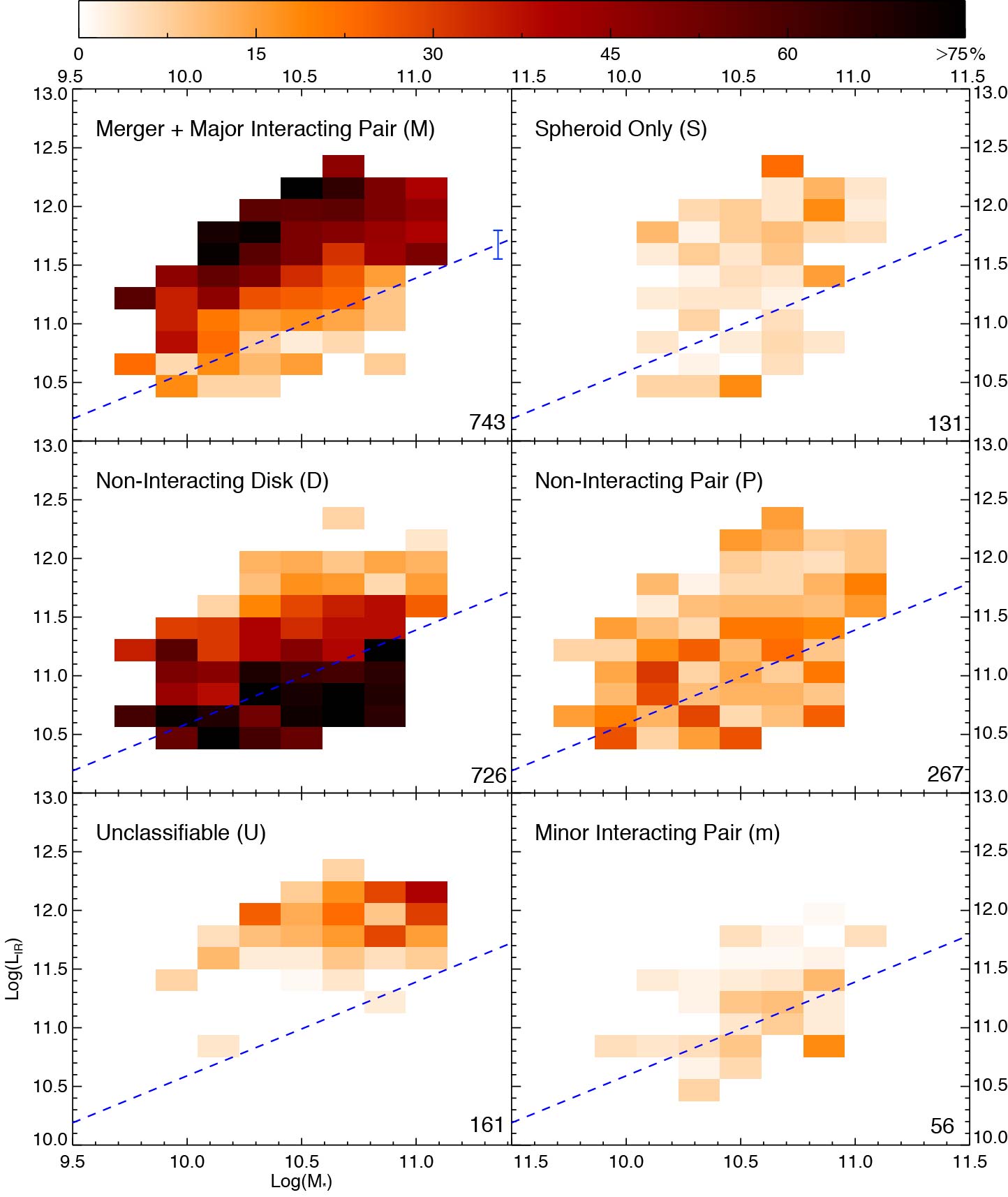}
\caption{
Six panels show the percentage of merger+major interacting pair, spheroid only, non-interacting disk, non-interacting pair, unclassifiable and minor interacting pair.
Each \LIR\ and \Mstar\ bin adds up to 100\% in all six panels.
The letters in the brackets correspond to the morphology classification results indicated in Fig.~\ref{fig:LIR_b4}$-$Fig.~\ref{fig:LIR_b1}.
The numbers in the bottom right corner of each panel indicate the numbers of galaxies of each category.
The increasing trend with \LIR\ seen in the top left panel shows similar information as the top panel in Fig.~\ref{fig:int_three} since the dependence of interaction fraction on \Mstar\ is small.
%The color gradient demonstrates a clear increasing trend of interacting systems with respect to increasing of \LIR. 
The blue dashed line in each panel is the SFR-\Mstar\ relation at the median redshift ($z=0.7$) of the entire sample, and the corresponding error bar in the top left panel indicates the dispersion of the SFR-\Mstar\ relation within $z=0.7\pm0.2$ redshift range.
} 
\label{fig:int_msplane}
\end{figure*}

In the top left panel of Fig.~\ref{fig:int_msplane}, we plot the percentage of interacting systems on the basis of \LIR\ and \Mstar.
This plot illustrates the correlations in Fig.~\ref{fig:int_three} in a two dimensional view.
The fraction of interacting systems increases fairly uniformly with \LIR\ at all observed \Mstar.
The rest of the panels in Fig.~\ref{fig:int_msplane} show results for the remainder of our objects that are not classified as interacting systems, including spheroid-only, non-interacting disk, non-interacting pair, minor interacting pair, and unclassifiable.
The sum of each \LIR\ and \Mstar\ bin adds up to 100\% over all six panels.
The distribution of non-interacting disks show a decreasing trend with \LIR, and no clear trends are seen in the other four panels.
The pure spheroids may be explained as late stage or the end product of mergers, and the non-interacting pairs may be the beginning stage of mergers.
However, since these categories do not show clear evidence of interaction, we choose not to combine them with interacting systems or non-interacting disks.

We have quantified the uncertainties in the interaction fraction when different criteria are used (the ambiguous flag).  
%For example, the presence of large-scale tidal features is a clear evidence of interactions \citep{Toomre1972, Barnes1992}.
%Irregular, clumpy structures without the presence of tidal features may also be explained as interactions with the dimming of tidal features.
%However, these structures may be alternatively created via secular accretion \citep{Dekel2009a}, where no large scale tidal features are created.
About 25\% of interacting galaxies are also assigned as ambiguous class, and they do not show distinct distribution compared to the entire interacting systems (Fig.~\ref{fig:int_three}).
The majority of candidates of high$-z$ clumpy disks lie well above the SFR-\Mstar\ relation as other unambiguously classified interacting systems.

\subsection{The Comparison between $F814W$, $F125W$ and $F160W$ Morphologies}
%At $z\sim1$ and higher, the $F814W$ band starts to trace rest-frame UV instead of optical morphologies.
Seventy-two {\it Herschel}-selected galaxies have WFC3 $F125W$- and $160W$-band images from CANDELS \citep{Grogin2011,
Koekemoer2011} (18 at $0.2<z<0.5$, 37 at $0.5<z<1.0$ and 17 at $1.0<z<1.5$), and thus we can compare the morphology in $F814W$, $F125W$ and $F160W$ of these  galaxies.
We see no obvious differences are seen between the $F125W$ and $F160W$ images, hereafter we only use the $F125W$ images for our discussion due to their better spatial resolution.

In the lowest redshift bin, all galaxies preserve the same classifications between $F814W$ and $F125W$ images except for 3 galaxies that show more prominent spheroid components at the longer wavelength.
At $0.5<z<1.0$, the overall fraction of disk features from these two bands is similar, but we see a higher percentage of spheroid features (16\% vs. 37\%), and a lower percentage of irregular features (40\% vs. 20\%) from the classifications based on the $F125W$ images.
In the highest redshift bin where the $F814W$ images probe rest-frame $U$-band, a higher percentage of disk features (30\% vs. 60\%) and lower percentage of irregular features (60\% vs. 35\%) are observed with $F125W$ images.
As for the interaction class, about 20\% $-$ 30\% of galaxies at $0.5<z<1.5$ have different classifications.
About half of these high redshift objects show less clumpy structure in the $F125W$ images, and they tend to be recognized as non-interacting galaxies (but we also notice that the lower resolution $F125W$ images compared to $F814W$ images can smooth out detailed structures).
In the other half, the secondary companion or large tidal features are only detected at longer wavelengths, thus increasing the interaction fraction.
In the highest redshift bin, two unclassifiable objects in $F814W$ images are both classified as interacting galaxies.

The differences seen between the $F814W$ and $F125W$ images for these 72 galaxies demonstrate the difficulty in studying galaxy morphology using only one band, particularly in between rest-frame UV and optical.
The morphological features can vary due to band-shifting, especially when identifying disk features at $z>1$ and irregular features at $z>0.5$.
However, we emphasize that this work has split the entire sample into three redshift bins which correspond to rest-frame $V$-, $B$-, and $U$-bands, and we focus on investigating how galaxy morphologies vary with their intrinsic properties (\LIR, \Mstar\ and the location on the SFR-\Mstar\ relation) in each redshift bin.
Within individual redshift bins, any morphological trend does not vary strongly with redshift.
The impact from band-shifting on our conclusions should only be significant if the band-shifting effects are strongly correlated with \LIR\ and \Mstar.

%In fact, the overall fraction of interacting systems may remain similar due to two competing factors: (1) less irregular features at longer wavelengths (owing to lower image resolution or galaxy intrinsic structure), and (2) more sensitive detections of tidal features and companion galaxies at longer wavelengths.

%\subsection{Galaxy Morphology in SFR-selected Sample}

\section{Discussion}
\subsection{The Role of Interaction in the SFR-\Mstar\ Relation}
By investigating morphological properties of 2084 {\it Herschel}-detected galaxies at $0.2<z<1.5$, we can address two important issues about high$-z$ star-forming and starburst galaxies:
(1) how do galaxy morphologies (which hint at the physical mechanisms that drive star formation) correlate with intrinsic galactic properties such as \LIR\ and \Mstar?
(2) what is the role of galaxy interactions in driving star formation and starburst activity at $z\sim1$?

In Fig.~\ref{fig:int_three}, the fraction of interactions shows clear increasing trends with both \LIR\ and the distance from the SFR-\Mstar\ relation.
Such an increasing trend with \LIR\ is consistent with the findings for (U)LIRGs in the local Universe \citep[e.g.][]{Ishida2004} and for the {\it Spitzer} 70 \mum-selected galaxies at $z\sim1$ \citep{Kartaltepe2010a}, in which interacting systems/mergers become dominant at higher \LIR.
The increasing trends with the distance from the SFR-\Mstar\ relations further demonstrate the growing role of galaxy interaction with increasing \LIR\ regardless of different stellar mass ($10^{10}\lesssim M_{*}\lesssim10^{11} M_{\odot}$).
The morphological analysis of local (U)LIRGs shows that merger stage is correlated with increasing \LIR, from separated galaxy pairs, strongly disturbed interacting systems, to more advanced mergers \citep[][Larson et al. in prep.]{Veilleux2002,Ishida2004,Haan2011}.
%Thus galaxy interactions are concluded as major drivers of enormous infrared emission and starbursts in local (U)LIRGs.
At $z\sim1$, we see the same increasing trends in the percentage of interacting systems with increasing \LIR\ (this work and K10), suggesting that mergers also play an important role in driving infrared luminosity at higher $z$.

Are galaxy interactions only important when galaxies fall well above the SFR-\Mstar\ relation in contrast to the normal star-forming, MS galaxies?
Based on the logarithmic distributions of $BzK$-selected and {\sc PACS}-selected galaxies at $1.5<z<2.5$, \citet{Rodighiero2011} conclude that only 2\% of star-forming galaxies represent high SFR outliers (i.e. starbursts, which are likely driven by major merger events) with respect to Gaussian-distributed, MS galaxies.
However, without morphology information, it remains unclear if all galaxies on the MS are evolving via secular accretion and only off-MS galaxies are interacting/merger systems.
In this work, we probe galaxy morphology on and well above the MS at $z\sim1$ based on 2084 {\it Herschel}-selected galaxies.
We determine the percentage of interacting systems along the MS is $\sim18\%$ ($\pm3\%$) given a typical dispersion of 0.3 dex in the SFR-\Mstar\ relation \citep[e.g.][]{Noeske2007} in comparison to the $\sim51\%$ non-interacting disk MS galaxies. 
The percentage remains significant ($\sim15\%$) even if we exclude galaxies with ambiguous classifications that can be explained as either clumpy disks or interacting galaxies.
In fact, this percentage can be even higher ($\sim25\%$) if some of the unclassifiable systems are interacting/merging galaxies.
This percentage (15\%$-$25\%) is similar to the findings at $z\sim2$, in which $\sim24\%$ ($\pm4\%$)  of MS galaxies ($\pm0.3$ dex) are interactions/mergers based on a sample of 122 (U)LIRGs \citep[][hereafter K12]{Kartaltepe2012}.
This interacting/MS population may evolve through a merger-driven evolutionary sequence \citep[e.g.][]{Mihos1996,Hopkins2006a} instead of following the evolutionary tracks predicted by secular gas exhaustion/regulation models \citep{Noeske2007a,Bouche2010}.

%\citet[][hereafter K12]{Kartaltepe2012} visually classify the morphology of 122 (U)LIRGs at $z\sim2$ and find $\sim24\%$ ($\pm4\%$) of MS galaxies ($\pm0.3$ dex) are interactions/mergers.
%The somewhat higher fraction in K12 may be due to the fact that the 122 (U)LIRGs mostly fall in the upper end of the MS yet our sample covers both ends of the MS, where the interaction fraction is lower when in the lower end of the MS.

\subsection{Comparisons with Morphological Analysis based on \sersic\ index}

\citet{Wuyts2011a} use \sersic\ index to characterize morphology for $\sim130,000$ galaxies at $z\sim1$, including galaxies in the COSMOS field.
They study galaxy morphology on the basis of SFR and \Mstar, and find a ``structural MS'' that has a \sersic\ index of $n\sim1$ along the MS and an increasing \sersic\ index as the distance from the MS increases \citep[$n=1$ gives the exponential profile and $n=4$ gives the de Vaucouleurs profile,][]{Sersic1963}.
The increasing trend of \sersic\ index with respect to the distance from the MS suggests that galaxies build up a central concentration of stellar mass as SFR increases at any \Mstar.
One possible physical mechanism to drive such transition of morphology is through galaxy interactions/mergers, where much of the gas content of the disk galaxies falls towards the center during galaxy encounters, and the violent relaxation re-distributes stars into a near de Vaucouleurs (\sersic\ index $n=4$) distribution \citep[e.g.][]{Toomre1972,Barnes1992}. 
This interpretation is consistent with the increasing fraction of interacting systems with respect to the distance from the MS seen the bottom panel of Fig.~\ref{fig:int_three}. 

The structural MS with a \sersic\ index of $n\sim1$ implies that the MS galaxies mainly show disk-like structures, and may be explained as ``normal'' star-forming galaxies.
This conclusion is somewhat contradictory compared to the significant fraction of interacting, MS galaxies identified in our work and K12.
However, we emphasize that the results based on both analyses are not necessary in disagreement.
This can be due to the difficulties in translating \sersic\ index and morphological type, particularly for interacting systems.
Although \sersic\ index may be a good proxy for disk galaxies ($n=1$)  and elliptical galaxies ($n=4$), it does not accurately represent interacting systems due to its parametric form \citep{Lotz2004}.
For example, the automatic \sersic\ index fitting algorithm often fails to identify interacting pairs.
Furthermore, galaxy interactions can result in messy and clumpy structures, which often yield a small \sersic\ index since the overall light distribution is even flatter than that in normal disks.
To illustrate this statement, we have measured the \sersic\ index of 36 local (U)LIRGs, all of which show clear signatures of interactions from GOALS-SDSS images \citep[The Great Observatories All-Sky LIRG Survey;][]{Armus2009}.
We find a median \sersic\ index of 1.8 for these 36 galaxies, which include all merger stages (Hung et al., in prep.).
Given the complications in translating \sersic\ index to galaxy morphology types, we favor conclusions based on visual classifications.

\section{Conclusions}

We have carried out a detailed morphological analysis of a large sample of 2084 {\it Herschel}-selected (PACS + SPIRE) galaxies with redshifts $0.2<z<1.5$ in the full 2-deg$^2$ COSMOS  field.
A detailed visual classification scheme is employed to classify each object  (e.g. disk, spheroid, irregular, strong interaction/merger).  
We then compare the distribution of morphological types with \LIR\ as well as distance above and below the $SFR$-\Mstar\ relation (``galaxy main sequence").
Our conclusions are summarized below:
\begin{enumerate}
\item The percentage of galaxies classified as ``disks"  decreases systematically with increasing \LIR\ at all values of \Mstar\ whereas the percentage of objects classified as ``irregular"  systematically increases with \LIR.
\item The percentage of strongly interacting/merger systems increases with  \LIR\ as well as distance above the SFR-\Mstar\ relation, with $\sim50\%$  of galaxies classified as interacting/merger systems at \LIR\ $>10^{11.5}$ \Lsun.
\item We also find that a significant percentage ($\gtrsim18\%$) of the infrared-luminous, main sequence galaxies ($\pm$0.3 dex) show signatures of strong interaction/mergers, suggesting that this population may not evolve through the evolutionary tracks predicted by simple gas exhaustion models.   

\end{enumerate}

\acknowledgments
C.-L. Hung thanks V. U and J. Chu for their help with visual classification at the early stage of this project. 
D. B. Sanders and C. M. Casey acknowledge the hospitality of the Aspen Center for Physics, which is supported by the National Science Foundation Grant No. PHY-1066293.
C. M. Casey is generously supported by a Hubble Fellowship from Space Telescope Science Institute, grant HST-HF-51268.01-A.

COSMOS is based on observations with the NASA/ESA {\em Hubble Space Telescope}, obtained at the Space Telescope Science Institute, which is operated by AURA Inc, under NASA contract NAS 5-26555; also based on data collected at : the Subaru Telescope, which is operated by the National Astronomical Observatory of Japan; the XMM-Newton, an ESA science mission with instruments and contributions directly funded by ESA Member States and NASA; the European Southern Observatory, Chile; Kitt Peak National Observatory, Cerro Tololo Inter-American Observatory, and the National Optical Astronomy Observatory, which are operated by the Association of Universities for Research in Astronomy, Inc. (AURA) under cooperative agreement with the National Science Foundation; the National Radio Astronomy Observatory which is a facility of the National Science Foundation operated under cooperative agreement by Associated Universities, Inc ; and the Canada-France-Hawaii Telescope operated by the National Research Council of Canada, the Centre National de la Recherche Scientifique de France and the University of Hawaii. 

PACS has been developed by a consortium of institutes led by MPE (Germany) and including UVIE (Austria); KU Leuven, CSL, IMEC (Belgium); CEA, LAM (France); MPIA (Germany); INAF-IFSI/OAA/OAP/OAT, LENS, SISSA (Italy); IAC (Spain). This development has been supported by the funding agencies BMVIT (Austria), ESA-PRODEX (Belgium), CEA/CNES (France), DLR (Germany), ASI/INAF (Italy), and CICYT/MCYT (Spain).
SPIRE has been developed by a consortium of institutes led by Cardiff University (UK) and including Univ. Lethbridge (Canada); NAOC (China); CEA, LAM (France); IFSI, Univ. Padua (Italy); IAC (Spain); Stockholm Observatory (Sweden); Imperial College London, RAL, UCL-MSSL, UKATC, Univ. Sussex (UK); and Caltech, JPL, NHSC, Univ. Colorado (USA). This development has been supported by national funding agencies: CSA (Canada); NAOC (China); CEA, CNES, CNRS (France); ASI (Italy); MCINN (Spain); SNSB (Sweden); STFC (UK); and NASA (USA).

\bibliographystyle{apj}
\bibliography{/Users/clhung/Documents/Cosmos}

\appendix
\section{A. Image gallery of {\it Herschel}-selected galaxies}
\begin{figure*}
 \centering
  \includegraphics[width=0.9\textwidth]{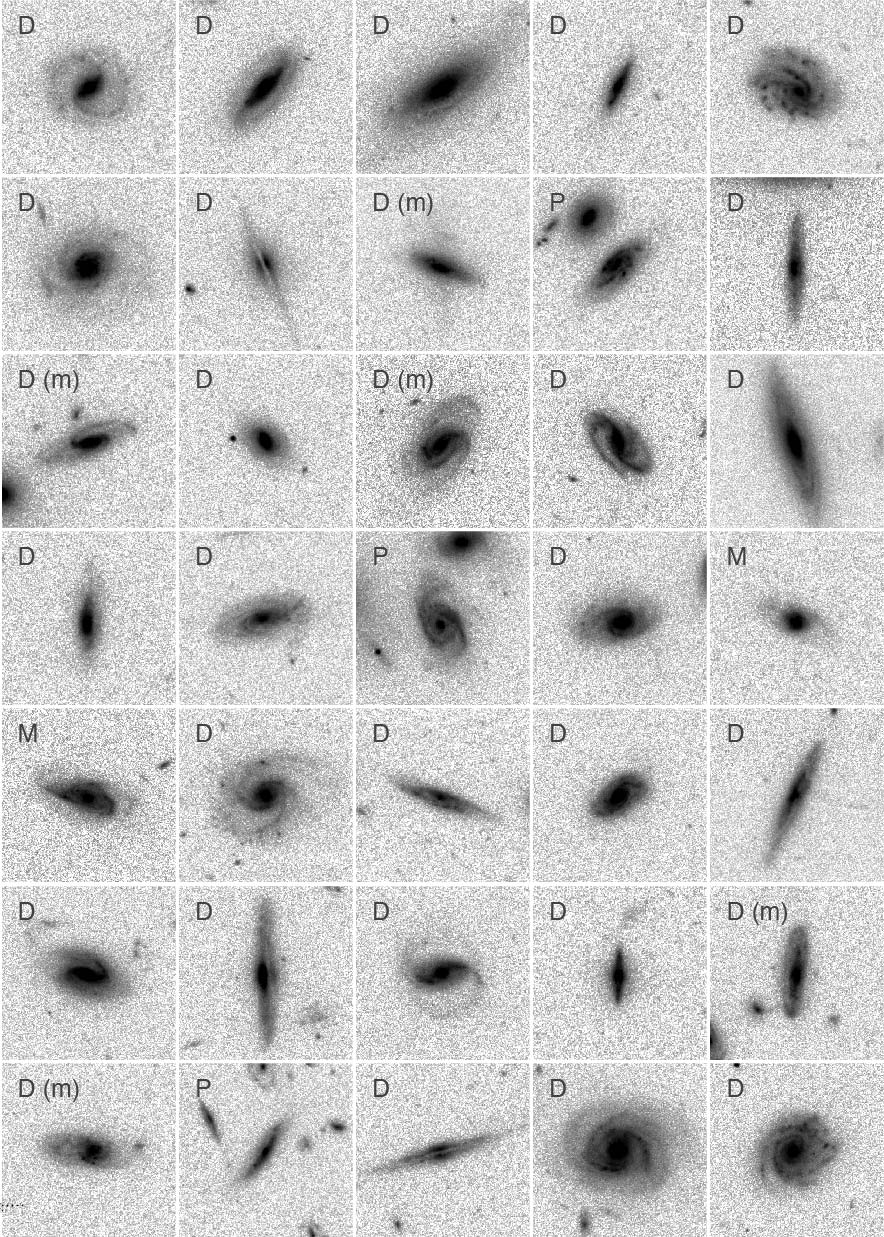}
\caption{
ACS $F814W$ band image gallery of randomly selected {\it Herschel}-selected galaxies at $0.37<z<0.45$ with \Mstar\ $>10^{10.5}M_{\odot}$ and \LIR\ $<10^{11}$\Lsun.
The length and width of each image correspond to a physical size of 50 kpc and an angular size of $\sim7$\arcsec at $z=0.7$.
The classification results (corresponding to the six panels in Fig.~\ref{fig:int_msplane} are indicated in the top-left corner (M: Merger or Major Interacting Pair, S: Spheroid Only, D: Non-Interacting Disk, P: Non-Interacting Pair, U: Unclassifiable, m: Minor Interacting Pair).
The possible alternative classifications are shown in brackets.
} 

\label{fig:LIR_b4}
\end{figure*}

\begin{figure*}
 \centering
  \includegraphics[width=0.9\textwidth]{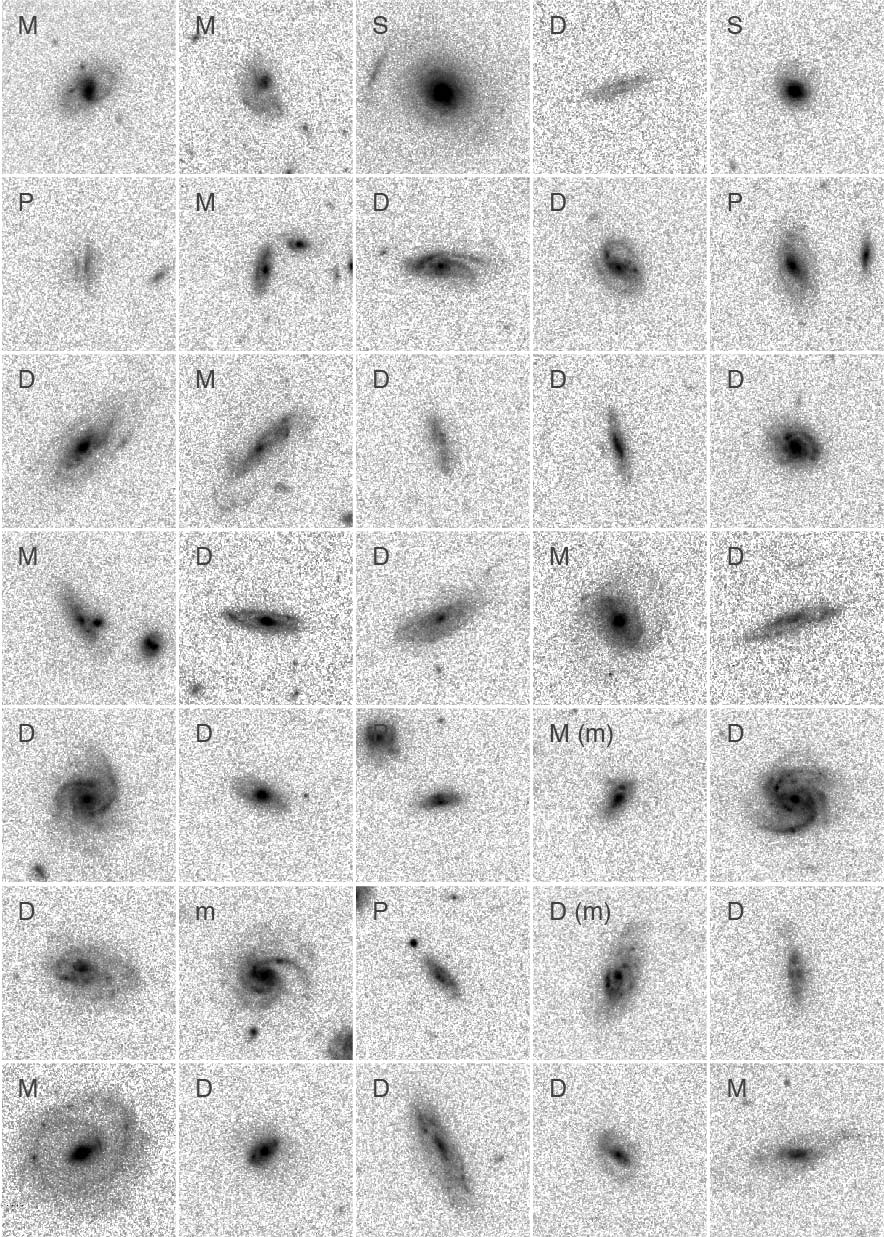}
%  \addtocounter{figure}{-1}
\caption{Same as Fig.~\ref{fig:LIR_b4}, galaxies at $0.65<z<0.72$ with \Mstar\ $>10^{10.5}M_{\odot}$ and $10^{11.0}<$\LIR$<10^{11.5}$\Lsun.
}
\label{fig:LIR_b3}

\end{figure*}

\begin{figure*}
 \centering
  \includegraphics[width=0.9\textwidth]{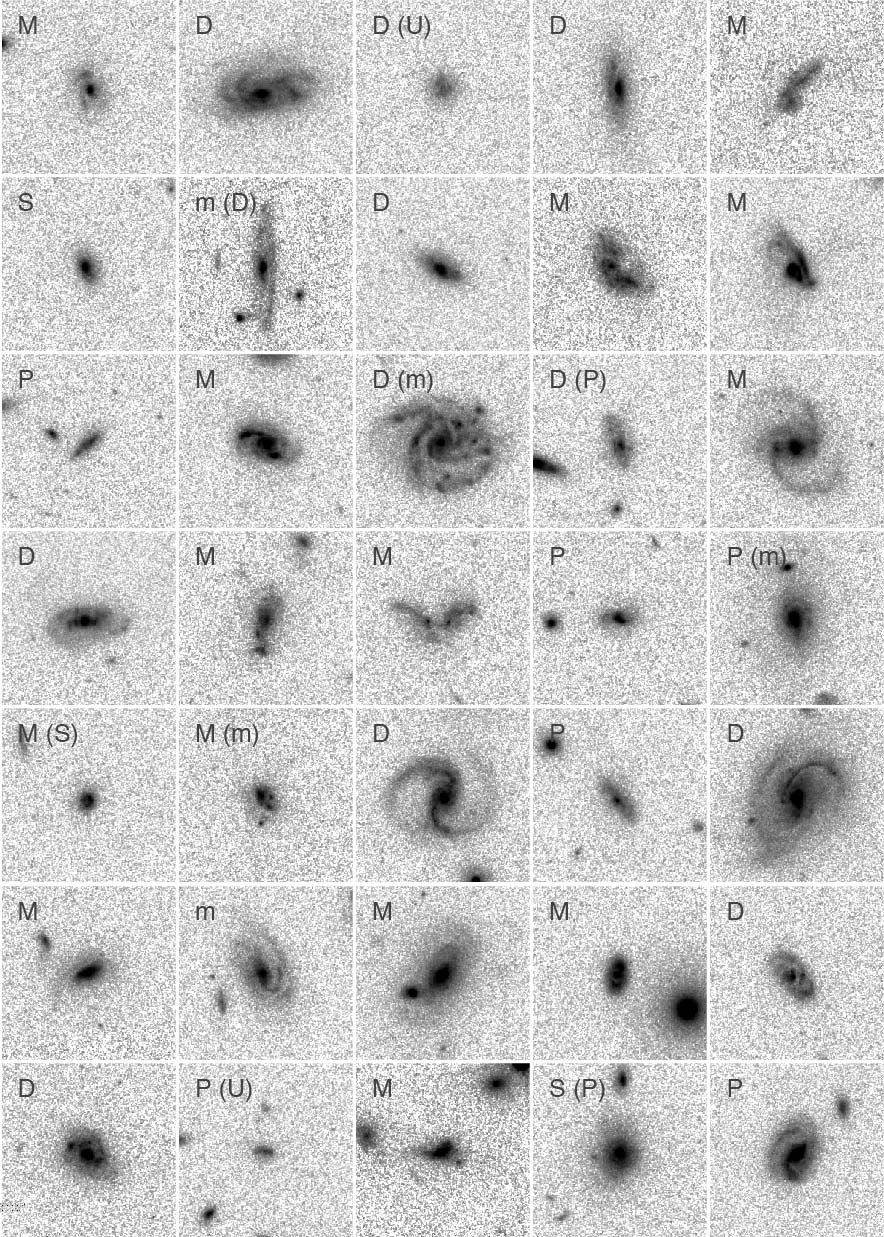}
%  \addtocounter{figure}{-1}
\caption{Same as Fig.~\ref{fig:LIR_b4}, galaxies at $0.67<z<0.72$ with \Mstar\ $>10^{10.5}M_{\odot}$ and $10^{11.5}<$\LIR$<10^{12}$\Lsun.
}
\label{fig:LIR_b2}
\end{figure*}

\begin{figure*}
 \centering
  \includegraphics[width=0.9\textwidth]{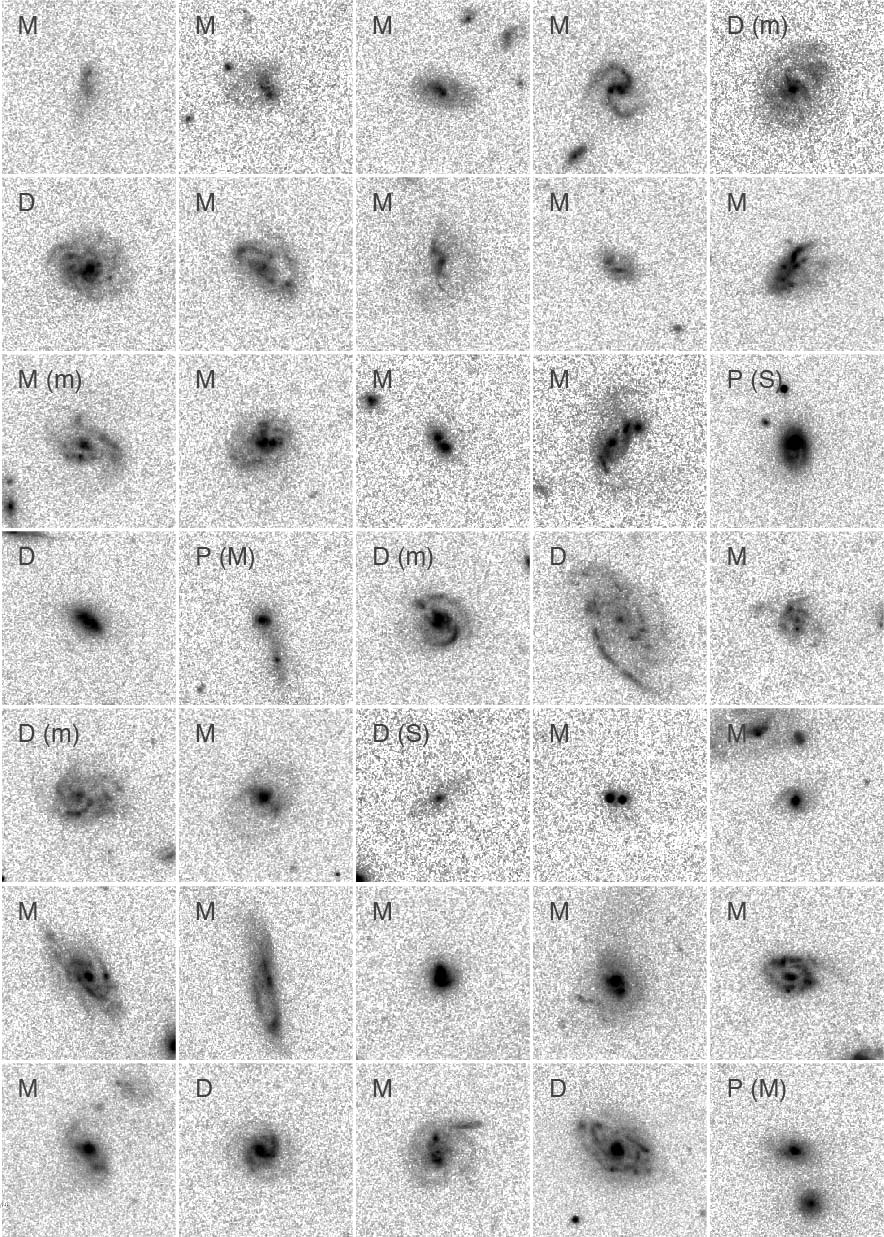}
\caption{Same as Fig.~\ref{fig:LIR_b4}, galaxies at $0.65<z<0.95$ with \Mstar\ $>10^{10.5}M_{\odot}$ and \LIR\ $>10^{12}$\Lsun.
} 
\label{fig:LIR_b1}
\end{figure*}

We present a subset of ACS images of {\it Herschel}-selected galaxies to demonstrate our visual classifications.
However, we note that adjusting the brightness and contrast of these images has been a necessary step in our classification processes, which cannot be demonstrated in these fixed-stretch images.
In Fig.~\ref{fig:LIR_b4} to Fig.~\ref{fig:LIR_b1}, we show a {\it randomly-selected} subset of galaxies with \Mstar$>10^{10.5}M_{\odot}$ in four \LIR\ bins.
Due to the strong dependence between $z$ and \LIR, as demonstrated in Fig.~\ref{fig:lirnum}, the four \LIR\ bins have been chosen in slightly different redshift range.

We indicate our visual classification results in the top-left corner of each image, with a possible alternative classification shown in brackets.
We find that more than 80\% of galaxies have clear conclusion of their classifications based on the discussion between three classifiers (CLH, DBS and JEB).
For those $\sim$17\% that may have alternative classifications, the confusions mostly originate in classifying non-interacting disks or minor interacting systems.

\end{document}